 \documentclass[useAMS,a4paper,fleqn,usenatbib]{mnras}
 
 \usepackage{times,graphicx,amssymb,amsmath,natbib}

\title[A simple method for solving chemical evolution]{A simple and general 
method for solving detailed chemical evolution with delayed production of iron and 
other chemical elements}
\author[F. Vincenzo et al.]{F. Vincenzo$^{1,2}$\thanks{E-mail:
vincenzo@oats.inaf.it}, F. Matteucci$^{1,2,3}$ and E. Spitoni$^{1}$
\\
$^{1}$Dipartimento di Fisica, Sezione di Astronomia, Universit\`a di Trieste, via G.B. Tiepolo 11, 34100, Trieste, Italy\\
$^{2}$INAF, Osservatorio Astronomico di Trieste, via G.B. Tiepolo 11, 34100, Trieste, Italy\\
$^{3}$INFN, Sezione di Trieste, Via Valerio 2, 34100, Trieste, Italy  
}

\begin{document}

\date{Accepted 2016 December 23. Received 2016 December 19; in original form 2016 June 22.}

\pagerange{\pageref{firstpage}--\pageref{lastpage}} \pubyear{2016}

\maketitle

\label{firstpage}


\begin{abstract}

\noindent We present a theoretical method for solving the chemical evolution of galaxies, by assuming an 
instantaneous recycling approximation for chemical elements restored by massive stars and the 
Delay Time Distribution formalism for the delayed chemical enrichment by Type Ia Supernovae.  
The galaxy gas mass assembly history, together with the assumed stellar yields and initial mass function, 
represent the starting point of this method. We derive a simple and general equation which closely relates 
the Laplace transforms of the galaxy gas accretion history and star formation history, which can be used 
to simplify the problem of retrieving these quantities in the galaxy evolution models assuming a linear 
Schmidt-Kennicutt law. We find that -- once the galaxy star formation history has been reconstructed 
from our assumptions -- the differential equation for the evolution of the chemical element $X$ can be suitably 
solved  with classical methods. We apply our model to reproduce the $[\text{O/Fe}]$ and $[\text{Si/Fe}]$ vs. $[\text{Fe/H}]$ 
chemical abundance patterns as observed at the solar neighborhood, by assuming a decaying exponential infall 
rate of gas and different delay time distributions for Type Ia Supernovae; we also explore the effect of assuming 
a nonlinear Schmidt-Kennicutt law, with the index of the power law being $k=1.4$. Although approximate, we 
conclude that our model with the single degenerate scenario for Type Ia Supernovae provides the best agreement 
with the observed set of data. Our method can be used by other complementary galaxy stellar population 
synthesis models to predict also the chemical evolution of galaxies. 

\end{abstract}



\section{Introduction} 

Understanding the evolution of the chemical abundances within the interstellar medium (ISM) of galaxies 
is fundamental for the development of a galaxy formation and evolution 
theory which aims at being complete. 
Numerical codes of chemical evolution address this issue; they 
can be thought of as a particular realization of stellar population synthesis models, 
constraining the evolution of galaxies from the perspective of their observed 
chemical abundance patterns. From a pure theoretical point of view, 
since the spectral and broad-band photometric properties of stars 
depend on their initial metallicity, it is clear that the galaxy chemical evolution should be understood 
first, before drawing the galaxy spectro-photometric evolution. 

Chemical evolution models compute the rate of restitution 
of a given chemical element $X$, at any time $t$ of galaxy evolution, 
by taking into account all the stars which die at that time. 
The latter quantity corresponds to an 
integral appearing in the differential equations of chemical evolution models, involving 
the galaxy star formation history (SFH), the initial mass function (IMF) and the nucleosynthetic stellar yields. 
On the other hand, spectro-photometric models recover the integrated spectrum of a galaxy 
at any time $t$ of its evolution, by taking into account all the stars which are still alive at that time. 
This quantity is computed by solving two folded integrals involving the galaxy SFH, the IMF and the stellar spectra. 

Given the fact that the integrals involved in chemical and spectro-photometric models are similar, 
their extremes are complementary: 
the integrated light from a galaxy is contributed by all the stars which are alive at that time; 
the chemical elements are mainly contributed (to a first approximation) by all the stars which are dying at that time.  
The difficulty of building up a fast and accurate chemical evolution model -- including  
chemical elements restored with a certain delay time from the star formation event -- 
to couple with other population synthesis models, has represented an obstacle for 
 developing a complete galaxy formation and evolution model for different groups in the past. 
 Examples of works combining a chemical evolution model with the galaxy 
 spectro-photometric evolution can be considered those of  
 \citet{brocato1990,bressan1994,gibson1997,silva1998,boissier2000,calura2008,cassara2015} 
 for unresolved galactic stellar systems and \citet{vincenzo2016b} for resolved stellar systems. 

In this work, we present a simple and fast method to solve chemical evolution of galaxies, 
by assuming an instantaneous recycling approximation 
(IRA)\footnote{The stellar lifetimes of the stars with mass $m\ge1\,\text{M}_{\sun}$ are neglected in the 
differential equations, while they are assumed to be infinite for stars with $m<1\,\text{M}_{\sun}$.} 
for chemical elements contributed by massive stars and the Delay Time Distribution (DTD) formalism 
for chemical elements restored by Type Ia Supernovae (SNe). We aim at reproducing the $[\alpha/\text{Fe}]$ vs. 
$[\text{Fe/H}]$ chemical abundance patterns as observed in the Milky Way (MW) at the solar neighborhood, by exploring the 
effect of different prescriptions for the DTD of Type Ia SNe. The assumed galaxy gas mass assembly history 
represent the starting point of this model. This can be easily incorporated in other 
stellar population synthesis models to better characterize, in a simple but effective way, the formation and evolution of 
galaxies from the observed properties of their stellar populations at the present time. 

Our work is organized as follows. In Section \ref{sec:model}, we present the theoretical framework of our model, 
namely the main assumptions  and the set of differential equations which we solve numerically. 
In Section \ref{sec:data}, we describe the observed set of data for the MW. 
In Section \ref{results}, we present the results of our model, which aims at reproducing the observed $[\text{O/Fe}]$ and 
$[\text{Si/Fe}]$ vs. $[\text{Fe/H}]$ chemical abundance patterns, 
as observed in the solar neighborhood. Finally, in Section \ref{conclusions}, we end with our conclusions.


\section{Theoretical framework} \label{sec:model} 

Analytical models of chemical evolution have been widely used in the past by theorists and observers to 
predict the metallicity evolution of a stellar system in a simplified but, at the same time, 
highly predictive fashion. 
In fact, although these models have intrinsic shortcomings, 
they provide a very good approximation for the chemical evolution of elements like oxygen, 
which are restored on short typical time scales from the star formation event. 
Incidentally, oxygen is also the dominant chemical species among the metals in the 
galaxy ISM, since it is mainly synthesized by massive stars, with lifetimes $\tau \lesssim30\, \text{Myr}$, 
dying as core-collapse SNe. 

Beyond the metallicity evolution, 
analytical models of chemical evolution are also able to
retrieve the evolution of the galaxy stellar and gas mass with time; therefore, 
they can draw an approximate but 
complete physical picture for the evolution of the various galaxy mass components. 
We remind the reader that the galaxy star formation history simply follows from the gas mass evolution, 
because of the Schmidt-Kennicutt law 
which is assumed in the models. 

Some analytical models do not take into account the fact that the majority of the chemical elements 
in the ISM usually have more than one nucleosynthesis channel; to complicate further this 
scenario, each nucleosynthesis channel is also characterized by a distinctive 
distribution of typical time scales for the chemical enrichment, which differs from the other. 

In this Section, we describe our method for solving the problem of coupling 
different nucleosynthesis channels in analytical chemical evolution models; 
in particular, we show how the nucleosynthesis from core-collapse SNe can be coupled with the one by Type Ia SNe in 
a simplified but effective theoretical picture. We think that this method can useful both for observers and theorists 
who wish to decouple -- for example -- the evolution of chemical elements like oxygen and iron, which do not trace each 
other and are usually used as tracers of the galaxy metallicity. 

\subsection{The delay time distribution of Type Ia supernovae}

Chemical elements like iron, silicon or calcium are also synthesized by Type Ia SNe beyond core-collapse Supernovae. 
It is therefore fundamental to properly take into account the Type Ia SN contribution in 
a theoretical model of chemical evolution. 
A useful formalism to compute the Type Ia SN rate in galaxies was developed by \citet{ruiz1998}, 
by originally introducing the concept of DTD in the theoretical framework (see also \citealt{strolger2004,greggio2005}). 
In this formalism, 
the Type Ia SN rate is defined as the convolution of the galaxy star formation rate (SFR) with a suitable DTD, as follows.   
\begin{equation} 
R_{\text{Ia}}(r,t) = C_{\text{Ia}} \int \limits_{\tau_{1}}^{\min(t,\tau_2)} d\tau \, \text{DTD}_{\text{Ia}}(\tau) \, \psi(r,t-\tau),  
\label{eq:Ia_DTD}
\end{equation}
where $\tau_1$  and $\tau_2$ are suitable values depending on the adopted scenario for the DTD. The quantity 
$\psi(t)$ represents the galaxy SFR, with the units $\text{M}_{\sun}\,\text{pc}^{-2}\,\text{Gyr}^{-1}$, and 
the normalization constant $C_{\text{Ia}}$ is related to the fraction of stars in the binary systems giving rise to Type Ia SNe. 
The function $\text{DTD}_{\text{Ia}}(\tau)$ physically represents the (unnormalized) number of Type Ia SNe, 
which are expected to explode at the time $t=\tau$ from a burst of star formation 
at $t=0$, per unit mass of the simple stellar population (SSP) and per unit time of duration of the burst.  

Observational evidence suggests an integrated number of Type Ia SNe, which is $\sim1$ SN per $M_{\star}=10^{3}\,\text{M}_{\sun}$ of 
stellar mass formed, with a scatter of about  $2$-$10$ around this value \citep{bell2003,maoz2014}. 
In this work, the normalization constants of the examined DTDs are chosen 
so as to fulfill this criterion; in particular, the constant $C_{\text{Ia}}$ is computed by requiring the following constraint: 
\begin{equation} \label{eq:norm_cost}
\frac{\int_{r_1}^{r_2}dr\,r \int_{0}^{t_{\text{G}}}dt'\,R_\text{Ia}(r,t')}{\int_{r_1}^{r_2}dr\,r\int_{0}^{t_{\text{G}}}dt'\,\psi(r,t')}
= \frac{2\,\text{SNe}}{10^3\,\text{M}_{\sun}},
\end{equation} 
where $r_1$ and $r_2$ are defined as the inner and outer radii, where the integrated number of Type Ia SNe (the numerator 
in the left-hand side of equation \ref{eq:norm_cost}) and the galaxy stellar mass formed (the denominator) are computed.

According to \citet{greggio2005}, the so-called ``double degenerate scenario" 
-- in which Type Ia SNe originate from binary systems 
of electron degenerate CO white dwarfs losing angular momentum via gravitational wave emission -- 
can be modeled by assuming  
$\text{DTD}_{\text{Ia}}(\tau)\propto1/\tau$ (see also \citealt{totani2008}), 
which provides 
almost the same final results for the Type Ia SN rate in galaxies as the DTD proposed by \citet{schonrich2009} 
and recently assumed by \citet{weinberg2016}. 

An other widely used DTD often assumed in the literature is the so-called ``bimodal DTD" 
as determined by \citet{mannucci2006}. It was originally defined with 
the following functional form:
\begin{equation} 
\text{DTD}_{\text{Ia}}(\tau) \propto A_{1}\exp\big({-\frac{(\tau - \tau')^{2}}{2\sigma'^{2}}}\big) 
    + A_{2}\exp( -\tau/\tau_{\text{D}}), 
\end{equation}
where the constants $A_{1}\approx19.95$ and $A_{2}\approx0.17$ guarantee that the two terms in the equation 
equally contribute by $50$ per cent; the parameters determining the prompt Gaussian function are $\tau'=0.05\,\text{Gyr}$ and 
$\sigma'=0.01\,\text{Gyr}$; finally, the time scale of the tardy declining exponential component is $\tau_{\text{D}}=3\,\text{Gyr}$.

In this work, we also test the scenario in which Type Ia SNe originate from the C-deflagration of an electron-degenerate 
CO white dwarf accreting material from a red giant or main sequence companion (the so-called ``single degenerate scenario"). 
We assume for the single degenerate scenario the same prescriptions as given in \citet{matteucci_recchi2001}.  

\subsection{The approximate differential equation for the evolution of a generic chemical element} 

In this work, we assume IRA for the chemical elements restored 
by massive stars and the DTD formalism for the delayed chemical enrichment by Type Ia SNe; 
in particular, we solve the following approximate differential equation for the evolution of 
the surface gas mass density of a generic chemical element $X$ within the galaxy ISM: 
\begin{equation} \label{eq:diff_equation}
 \frac{d\sigma_{X}(r,t)}{dt} =  - X(r,t)\,\psi(r,t)\, + \hat{\mathcal{E}}_{X}(r,t)  
   + \hat{\mathcal{O}}_{X}(r,t) + \hat{\mathcal{R}}_{X,\text{Ia}}(r,t)   
\end{equation}
where we assume that the infall gas is of primordial chemical composition. 
The physical meaning of the various terms 
in the right-hand side of equation (\ref{eq:diff_equation}) can be summarized as follows. 
\begin{enumerate}
\item The first term represents the surface gas mass density of $X$  
which is removed 
per unit time and surface from the galaxy ISM because of the star formation activity. 
The quantity $X(r,t) = \sigma_{X}(r,t)/\sigma_{\text{gas}}(r,t)$ is the abundance by mass of the 
chemical element $X$. 
\item The second term, $\hat{\mathcal{E}}_{X}(r,t)$, takes into account both the stellar contributions to the 
enrichment of the newly formed chemical element $X$ and the
unprocessed quantity of $X$, returned per unit time and surface by dying stars 
without undergoing any nuclear processing in the stellar interiors. 
By assuming IRA, one can easily demonstrate that this term can be approximated as follows \citep{maeder1992}: 
\begin{equation}
\hat{\mathcal{E}}_{X}(r,t) =  X(r,t)\,\psi(r,t)\,R + \left< y_{X} \right> \big( 1-R \big)\, \psi(r,t), 
\end{equation} where the quantity $R$ represents the so-called ``return mass fraction", 
namely the total mass of gas returned into the ISM by a SSP, per unit mass of the SSP 
(see also \citealt{calura2014} for a more accurate approximation to compute the gas mass returned 
by multiple stellar populations), and $\left< y_{X}  \right>$ is the net yield of $X$ per stellar generation 
\citep{tinsley1980}. 
\item The third term in equation (\ref{eq:diff_equation}) removes the quantity of the chemical element $X$ 
which is expelled out of the galaxy potential well because of galactic winds. We assume the galactic wind 
to be always active over the whole galaxy lifetime and its intensity is 
directly proportional to the galaxy SFR, namely 
\begin{equation}
\hat{O}_{X}(r,t)=\omega\,\psi(r,t), 
\end{equation} 
\noindent with $\omega$ being the so-called ``mass loading factor'', a free parameter in the models. 
We can also think at the galactic wind as a continuous feedback effect of the SFR, which warms up the galaxy ISM through 
stellar winds and Supernova explosions; 
as a consequence of this, the gas can move from the cold to the hot phase of the ISM and hence it 
might not be immediately available for further reprocessing by star formation. 

\item The forth term represents the amount of $X$ restored per unit time and surface by Type Ia SNe, where 
$\left< m_{X,\text{Ia}}  \right>$ is the average amount of $X$ synthesized by each single Type Ia SN event. 
In particular, 
\begin{equation} 
\hat{\mathcal{R}}_{X,\text{ia}}(r,t) = \left< m_{X,\text{Ia}}  \right> \,R_{\text{Ia}}(r,t),
\end{equation} 
\noindent with $R_{\text{Ia}}(r,t)$ being the Type Ia SN rate, as defined in equation (\ref{eq:Ia_DTD}). 
\end{enumerate}

\noindent In summary, by specifying all the various terms, we can rewrite equation (\ref{eq:diff_equation}) as follows: 
\begin{eqnarray} \label{eq:diff_equation2}
 \frac{d\sigma_{X}(r,t)}{dt} =  - \sigma_{X}(r,t)\,\frac{\psi(r,t)}{\sigma_{\text{gas}}(r,t)}\,\big( 1 + \omega - R  \big) \, + &  \nonumber \\
  + \; \left< y_{X} \right> \big( 1-R \big)\, \psi(r,t) + \left< m_{X,\text{Ia}}  \right> R_{\text{Ia}}(r,t). & 
\end{eqnarray}

\noindent We remark on the fact that equation (\ref{eq:diff_equation2}) can be numerically solved with classical methods 
(e.g., with the Runge-Kutta algorithm), once the star formation history (SFH) of the galaxy has been previously determined, 
either observationally or theoretically; in Section \ref{sec:SFH}, we show how we derive this information.  
In this work, we solve equation (\ref{eq:diff_equation2}) for oxygen, silicon and iron, 
by assuming $\left<y_{\text{O}}\right>=1.022\times10^{-2}$, $\left<y_{\text{Si}}\right>=8.5\times10^{-4}$, 
$\left<y_{\text{Fe}}\right>=5.6\times10^{-4}$ and $R=0.285$, which 
can be obtained by assuming the \citet{kroupa1993} IMF (see also \citealt{vincenzo2016}). 
We adopt the so-called $3/8$-rule forth-order Runge-Kutta method to solve equation (\ref{eq:diff_equation2}). 
Finally, we assume the stellar yields of Type Ia SNe from \citet{iwamoto1999}.

\subsection{The galaxy star formation history} \label{sec:SFH}

We test the effect of two distinct phenomenological laws for the SFR: i) a linear Schmidt-Kennicutt law, as assumed 
in some analytical and numerical codes of chemical evolution, and ii) a nonlinear 
Schmidt-Kennicutt law. In the first case, the differential equation for the evolution of the galaxy 
surface gas mass density is linear and it can be easily solved with the Laplace transform method, for any assumed 
smooth gas mass assembly history, acting in the equation as a source term. 
In the second case, since 
the differential equation is nonlinear in the unknown, a numerical technique is adopted to solve the problem of 
retrieving the galaxy SFH. 

\subsubsection{Linear Schmidt-Kennicutt law for the galaxy SFR}

In the case of a linear Schmidt-Kennicutt law, the galaxy SFR is assumed to be directly proportional to the galaxy 
gas mass density, namely 
\begin{equation} \label{eq:linear_SFR}
\psi(r,t) = \nu_{\text{L}}\,\sigma_{\text{gas}}(r,t), 
\end{equation} where $\nu_{\text{L}}$ is the so-called ``star formation efficiency'', a free parameter in the models, having the units 
of $\text{Gyr}^{-1}$. The galaxy SFH, namely the evolution of the galaxy SFR with time and radius, 
can be theoretically recovered under IRA, by solving the following approximate differential equation for the evolution of the 
galaxy surface gas mass density: 
\begin{equation} \label{eq:gas_mass}
\frac{d\sigma_{\text{gas}}(r,t)}{dt} = - \psi(r,t)\,\big( 1 + \omega - R  \big) + \hat{\mathcal{I}}(r,t). 
\end{equation}  
The last term in equation (\ref{eq:gas_mass}) 
corresponds to the assumed galaxy gas mass assembly history, namely $\hat{\mathcal{I}}(r,t)= (d\sigma_{\text{gas}}/dt)_{\text{inf}}$. 
Most of the chemical evolution models in the literature customarily assume a decaying exponential infall rate of gas 
with time. Historically, the first work in the literature suggesting such an infall law is the one by 
\citet{chiosi1980}. Nevertheless, equation (\ref{eq:gas_mass}) can be numerically solved for any assumed 
galaxy gas accretion history with standard techniques (e.g., a Runge-Kutta algorithm).

 \begin{figure}
\includegraphics[width=9.3cm]{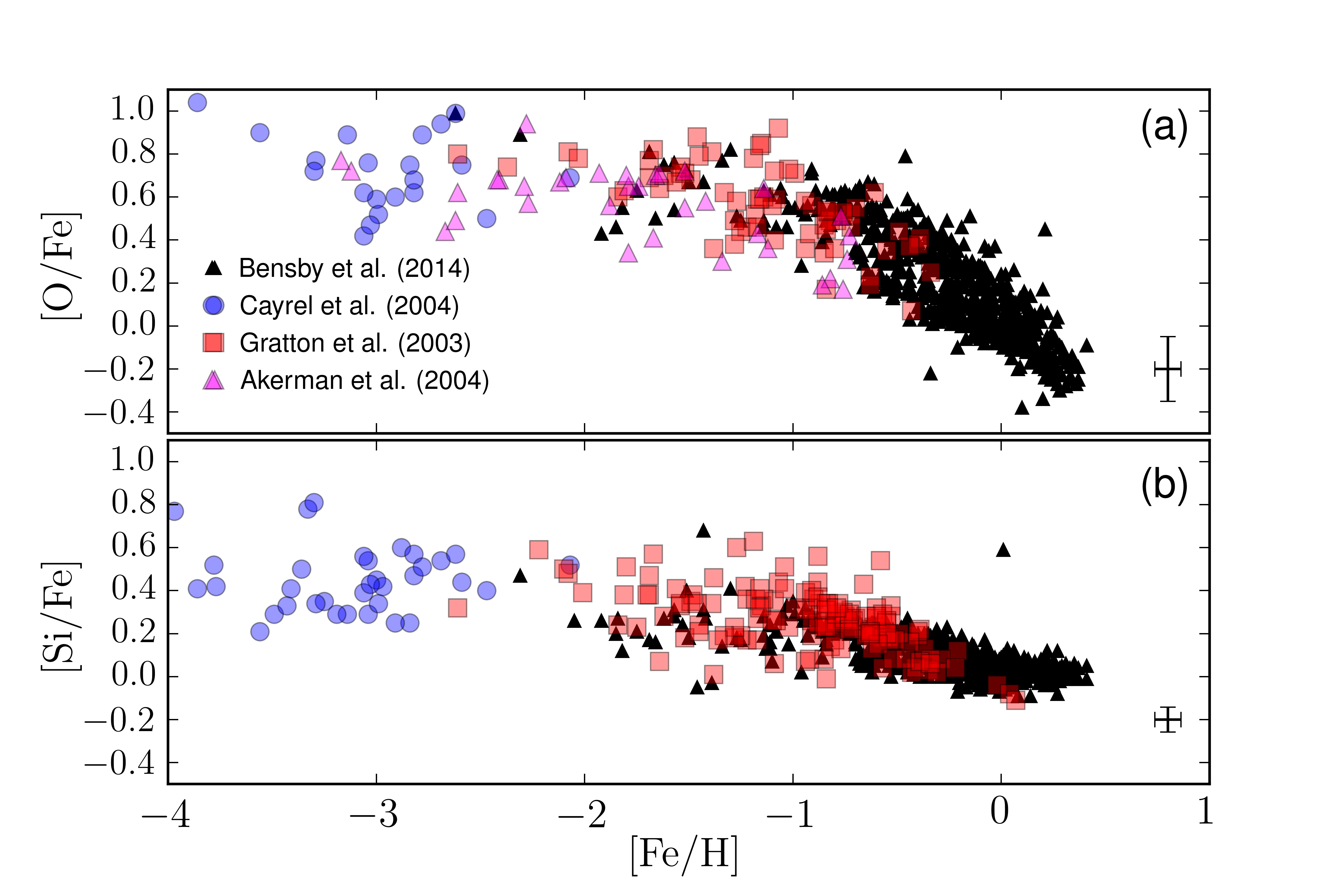}
\caption{ In this Figure, we show the observed data set for the $[\text{O/Fe}]$ (upper panel) and 
$[\text{Si/Fe}]$ (bottom panel) vs. $[\text{Fe/H}]$ chemical abundance patterns, that we assume in this work 
for the comparison with the predictions of our models. Data with different 
colors correspond to different works in the literature. In particular, our set of data for the 
solar neighborhood include MW halo 
stars from \citet[red squares]{gratton2003}, \citet[blue circles]{cayrel2004} and \citet[magenta triangles]{akerman2004}, 
and halo, thin and thick disc stars from \citet[black triangles]{bensby2014}. 
}
     \label{fig:data_by_author}
\end{figure} 

If we compute the Laplace transform of equation (\ref{eq:gas_mass}), it is straightforward to verify that 
the following equation can be obtained: 
\begin{equation} \label{laplace}
\mathcal{L}\big( \hat{\mathcal{I}}(r,t) \big)(s) = \frac{s+\alpha}{\nu_{\text{L}}}\,\mathcal{L} \big( \psi(r,t) \big) (s) - \sigma_{\text{gas}}(r,0),
\end{equation}
where $s$ is the frequency, with the units of $\text{Gyr}^{-1}$, and $\alpha= (1 + \omega - R)\,\nu$. 
Equation (\ref{laplace}) is very general 
and closely relates the Laplace transform of the galaxy gas accretion history, $\mathcal{L}\big( I(r,t) \big)$(s), with the Laplace 
transform of the galaxy SFH, $\mathcal{L}\big( \psi(r,t) \big)$(s), provided the SFR follows a linear Schmidt-Kennicutt law. 
Equation (\ref{laplace}) can be used both to retrieve the 
galaxy SFH from the assumed gas infall law and to solve the corresponding inverse problem, namely 
to reconstruct the galaxy gas mass assembly history from the 
observed SFH (one can assume a fitting function for the galaxy SFH to insert in equation \ref{laplace}). 

By assuming in equation (\ref{laplace}) the following law for the galaxy gas mass assembly history:
\begin{equation} \label{eq:infall}
\hat{\mathcal{I}}(r,t) = \sum_{j}{ \hat{\mathcal{I}}_{j}(r,t) } = \sum_{j}{ A_{j}(r) \, e^{-(t-t_{j})/\tau_{j}} \, \Theta(t-t_{j}) },
\end{equation}
which corresponds to a summation of separate gas accretion episodes, each obeying a decaying exponential 
law with time scale $\tau_{j}$ and starting at the time $t_{j}$, 
then it is straightforward to verify that the solution for the galaxy SFH is the following:
\begin{eqnarray} \label{eq:sfr_sol}
& \psi(r,t) =  \sum_{j}{ \frac{\nu_{\text{L}}\,A_{j}(r)}{\alpha - \frac{1}{\tau_{j}}}\,\Theta(t-t_{j})\,\Big[ e^{-(t-t_{j})/\tau_{j}} - e^{-\alpha(t-t_{j})} \Big]   } + \nonumber \\
      & + \, \nu_{\text{L}}\,\sigma_{\text{gas}}(r,0)\,e^{-\alpha t}\,\Theta(t). 
\end{eqnarray}
We remind the reader that the function $\Theta(t)$ in the equations above is defined as the Heaviside step function. 
Finally, the parameter $A_{j}(r)$ is a normalization constant, which fixes the total gas mass accreted 
by the $j$th accretion episode. 

For the sake of simplicity, in this work we assume the galaxy gas accretion history, as defined in equation (\ref{eq:infall}), 
to be composed of a single episode, starting at $t=0$. 
It can be shown that the functional form for galaxy SFH has the same expression 
as given in \citet{spitoni2016}. 

It is worth remarking that a linear relation is inferred in galaxies only between the galaxy SFR and the molecular 
hydrogen surface mass density, $\sigma_{\text{\textsc{H}}_2}$ \citep{leroy2008,bigiel2008,bigiel2011,schruba2011}, although 
the systematic uncertainties in this kind of studies might be still very large to draw firm conclusions.  Nevertheless, 
\citet{kennicutt1998} noticed that the relation between the inferred SFR and the gas mass density, $\sigma_{\text{gas}}$, 
can be well fit with a linear law, such as equation (\ref{eq:linear_SFR}), 
when the star formation efficiency is defined as the inverse of the typical dynamical time scale of the system, 
following the results of a previous theoretical study by \citet{silk1997}. 

\begin{figure}
\includegraphics[width=9.3cm]{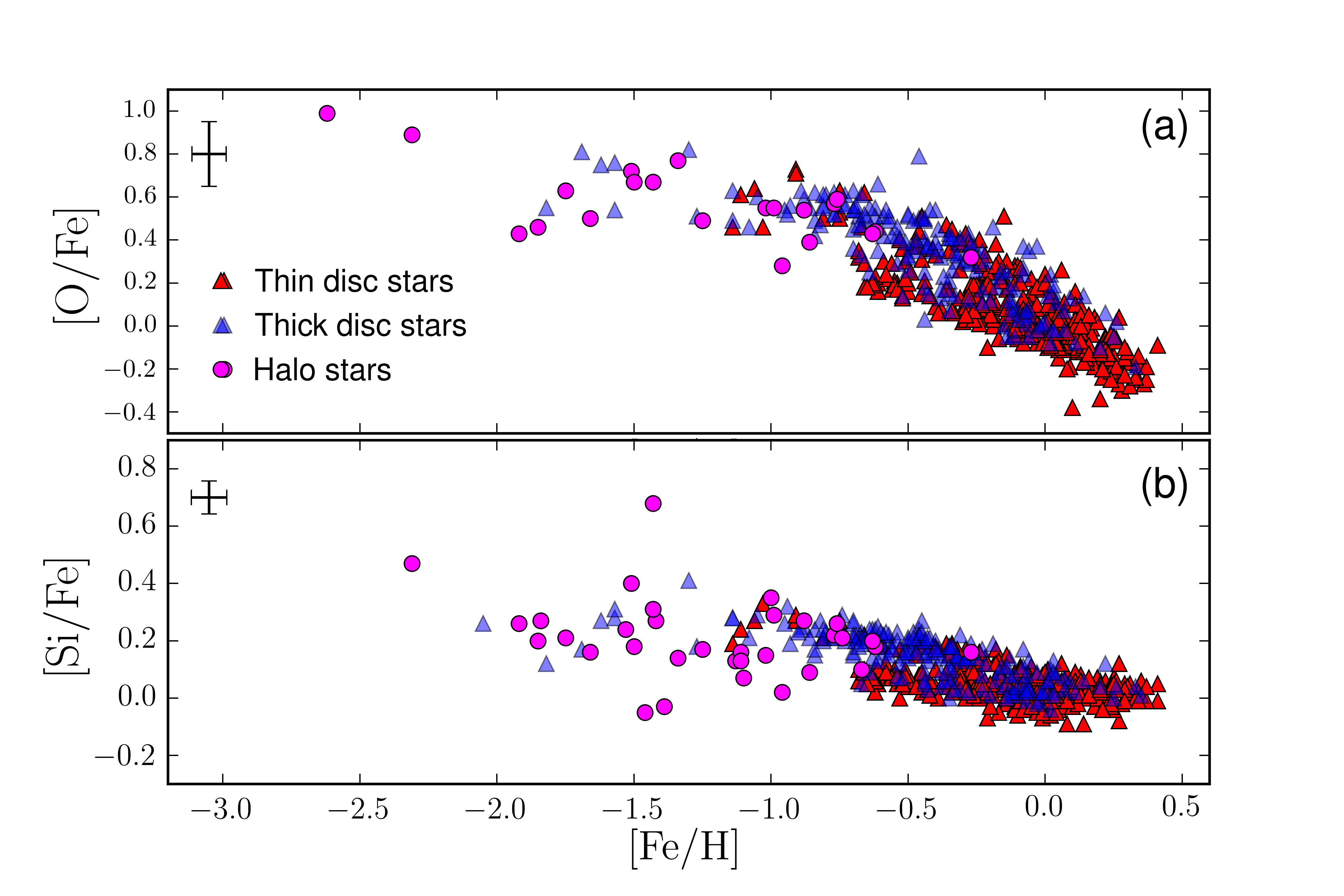}
\caption{ In this Figure, we show how the stars at the solar neighborhood in the \citet{bensby2014} sample are distributed 
among the various MW stellar components. Halo stars are drawn as magenta circles; thick disc stars 
as blue triangles; finally, thin disc stars are shown as blue triangles. We remind the reader that 
\citet{bensby2014} retrieve the membership of the stars in their sample with a kinematical criterion. 
}
     \label{fig:data_by_component}
\end{figure}

\subsubsection{Nonlinear Schmidt-Kennicutt law for the galaxy SFR} 

In this work, we also test the effect of assuming a nonlinear Schmidt-Kennicutt law for the evolution of the galaxy SFR, which 
has the following form: 
\begin{equation} \label{eq:original_kennicutt_law}
\psi(r,t) = \text{SFR}_{0}\,\left( \frac{\sigma_{\text{gas}}(r,t)}{\sigma_{0}} \right)^{k}, 
\end{equation} where, by following \citet{kennicutt1998}, we assume $k = 1.4$ and 
$\sigma_0 = 1\,\text{M}_{\sun}\,\text{pc}^{-2}$; 
finally, the quantity $\text{SFR}_{0}$ in equation (\ref{eq:original_kennicutt_law}) 
has the units of a star formation rate, namely 
$\text{M}_{\sun}\,\text{pc}^{-2}\,\text{Gyr}^{-1}$ (see also the discussion below). 
In our work, we treat the quantity $\text{SFR}_{0}$ as 
a free parameter, to reproduce the MW chemical abundance patterns. 

In the case of a nonlinear Schmidt-Kennicutt law, the star formation efficiency 
varies with radius and time, according to the following equation: 
\begin{equation} \label{eq:sfe_nonlinear}
\text{SFE}(r,t) = 
\frac{\psi(r,t)}{\sigma_{\text{gas}}(r,t)} = \frac{\text{SFR}_{0}}{\sigma_{0}}\,\left( \frac{\sigma_{\text{gas}}(r,t)}{\sigma_{0}} \right)^{k-1}. 
\end{equation} 

We remark on the fact that, in the original work by 
\citet{kennicutt1998}, 
$\text{SFR}_{0,\text{K98}}\approx2.5\times10^{-1}\,\text{M}_{\sun}\,\text{pc}^{-2}\,\text{Gyr}^{-1}$, 
which represents an average quantity, derived by fitting the observed relation between the SFR 
and $\sigma_{\text{gas}}$ in a sample of nearby star forming galaxies, which exhibit a spread in 
the $\text{SFR}$-$\sigma_{\text{gas}}$ diagram. 

In the physical picture drawn by equation (\ref{eq:original_kennicutt_law}), 
we can argue that -- as a consequence of a prolongated and continuous star formation activity, 
 which warms up the galaxy ISM through stellar winds and Type 
II SNe -- the galactic disc likely responds by regulating its dimensions (and hence its global thermodynamical quantities) 
to saturate towards a level of star formation, which is driven by the quantity $\text{SFR}_{0}$.  
In summary, the slope and power law index in equation (\ref{eq:original_kennicutt_law}) 
might represent ``\textit{truly basic 
physical constants}'', as pointed out by \citet{talbot1975}  
almost $40\,\text{years}$ ago about 
the SFR law in the galaxy formation and evolution models. 


We are aware that our considerations above are heuristic; in particular, 
there should be an underlying physical mechanism, 
common to almost all actively star forming stellar systems, 
which is not explained by our simple phenomenological recipes for star formation. 
A detailed physical theory must be developed, 
involving -- for example -- physical quantities in a statistical mechanics framework. 
In particular, one should be able to relate 
the astronomical quantity ``star formation rate'' with much more 
physical quantities, in order to gain a deeper knowledge about how 
a small scale phenomenon like the star formation can be regulated by large scale processes, and viceversa. 
An interesting attempt to develop a theory for the star formation activity in galaxies can be found in the work by 
\citet{silk1997}.


In principle, a possible method to solve equation (\ref{eq:original_kennicutt_law}) would be to use the Green functions. 
In particular, by assuming a nonlinear Schmidt-Kennicutt law, equation (\ref{eq:original_kennicutt_law}) 
becomes a Bernoulli differential equation with a source function, which is
given by the infall term, $\hat{\mathcal{I}}(r,t)$; 
the corresponding Green function is determined by the response of the system to a Dirac delta function 
in the infall term, which -- as aforementioned -- represents the source function in equation (\ref{eq:original_kennicutt_law});  
hence, from a physical point of view, the Green function corresponds to the solution of the so-called 
``closed-box model'', with all the gas being already present in the galaxy since the beginning of its evolution. 

The exact solution for $\sigma_{\text{gas}}$ with a nonlinear Schmidt-Kennicutt law could then be
found by convolving the Green function (i.e. the evolution of $\sigma_{\text{gas}}$ in a
closed box model, where the equation becomes a Bernoulli differential equation, with no source
term) with the assumed gas infall history. 

It is worth noting that, if $k=2$ in equation (\ref{eq:original_kennicutt_law}), as expected in 
the low density star-forming regions where almost all the gas is in its atomic form \citep{genzel2010}, 
then equation (\ref{eq:gas_mass}) for the evolution 
of $\sigma_{\text{gas}}$ with time can be put in the form of a Riccati equation, which can also be solved analytically. 


In summary, we firstly derive the evolution of $\sigma_{\text{gas}}$ with time by solving equation (\ref{eq:gas_mass}) with 
a numerical algorithm; then, it is straightforward to 
recover the galaxy SFR by making use of equation (\ref{eq:original_kennicutt_law}).

\subsection{Free parameters and methods} \label{subsec:free_par}


The free parameters of our model are given by 
i) the star formation efficiency, $\nu_{\text{L}}$, when the linear Schmidt-Kennicutt law is assumed, or 
the quantity $\text{SFR}_{0}$, in the case of a nonlinear Schmidt-Kennicutt law; 
ii) the mass loading factor, $\omega$, 
which determines the intensity of the galactic winds, and 
iii) the infall time scale, $\tau$, which characterizes the intensity of the gas infall rate, assumed to be a decaying exponential 
law.

The assumed initial mass function and the set of stellar 
yields determine the return mass fraction, $R$, the 
net yield of the chemical element $X$ per stellar generation, $\left< y_{X} \right>$, 
and the yield of $X$ from Type Ia SNe, $\left< m_{X,\text{Ia}}  \right>$. 
Other fundamental assumptions in the model are given by the DTD for the Type Ia SN rate and the prescription for the 
SFR law.

An observational constraint for the calibration of the gas infall history is given by the radial profile of the 
present-day total surface mass density, $\sigma_{\text{tot}}(r,t_{\text{G}})$, which is very difficult 
 to retrieve from an observational point of view. 
In this work, we apply our model to reproduce the observed chemical abundance patterns at the solar neighborhood, 
where $r_{\text{S}} = 8\,\text{kpc}$; 
hence we normalize the infall law, as given in equation (\ref{eq:infall}), 
by requiring the following constraint: 
\begin{equation}
\int\limits_{0}^{t_{\text{G}}}dt\,\hat{\mathcal{I}}(r_{\text{S}} , t) = \sigma_{\text{tot}}(r_{\text{S}}, t_{\text{G}}), 
\end{equation}
where we assume $t_{\text{G}}=14\,\text{Gyr}$, a single infall episode, 
and $\sigma_{\text{tot}}(r_{\text{S}} , t_{\text{G}}) = 54\,\text{M}_{\sun}\,\text{pc}^{-2}$, which 
represents the present-day total surface mass density at the solar neighborhood (see also \citealt{micali2013}, which 
refer to the work by \citealt{kuijken1991}).

The best model is defined as the one capable of reproducing the observed trend of the 
$[\text{O/Fe}]$ vs. $[\text{Fe/H}]$ chemical abundance pattern, which represents the best 
observational constraint for the chemical evolution models of the MW. 
Our best models are characterized by the following values of the free parameters: 
\begin{enumerate}
\item $\nu_{\text{L}}=2\,\text{Gyr}^{-1}$, for our best model with the linear Schmidt-Kennicutt law, 
and $\text{SFR}_{0}=2\,\text{M}_{\sun}\,\text{pc}^{-2}\,\text{Gyr}^{-1}$, for our best model with the nonlinear Schmidt-Kennicutt law; 
\item the infall time scale for the gas mass growth of the MW disc is  $\tau=7\,\text{Gyr}$; 
\item the mass loading factor is $\omega=0.4$. 
\end{enumerate} 
\noindent Finally, the best models assume the single degenerate scenario for the Type Ia SN rate. 
We remark on the fact that the predicted $[\text{Fe/H}]$-$\text{[O/Fe]}$ relation in the 
MW is fit by construction using fixed values for the key free parameters 
$\nu_{\text{L}}$ (or $\text{SFR}_{0}$), $\omega$ and $\tau$. 

It is worth noting that the values of our best parameters are 
in agreement with previous recent studies (see, for example, \citealt{minchev2013,nidever2014,spitoni2015}); 
in particular, a typical infall time scale $\tau\sim7\,\text{Gyr}$ and star formation efficiency 
$\sim1\,\text{Gyr}^{-1}$ are necessary to reproduce also the observed age-metallicity relation and the 
metallicity distribution function of the G-dwarf stars in the solar neighborhood \citep{matteucci2001,matteucci2012,pagel2009}. 

Once the best set of free parameters is determined, we fix them and 
explore the effects of assuming different DTDs for the Type Ia SN rate and different prescriptions 
for the relation between the SFR and the surface gas mass density, $\sigma_{\text{gas}}(r,t)$. In this way, 
we can demonstrate the flexibility of our method for solving the chemical evolution of galaxies, by taking 
into account the major systematics in the theory.

\section{The observed data set} \label{sec:data}

\begin{figure*}
\includegraphics[width=0.8\textwidth]{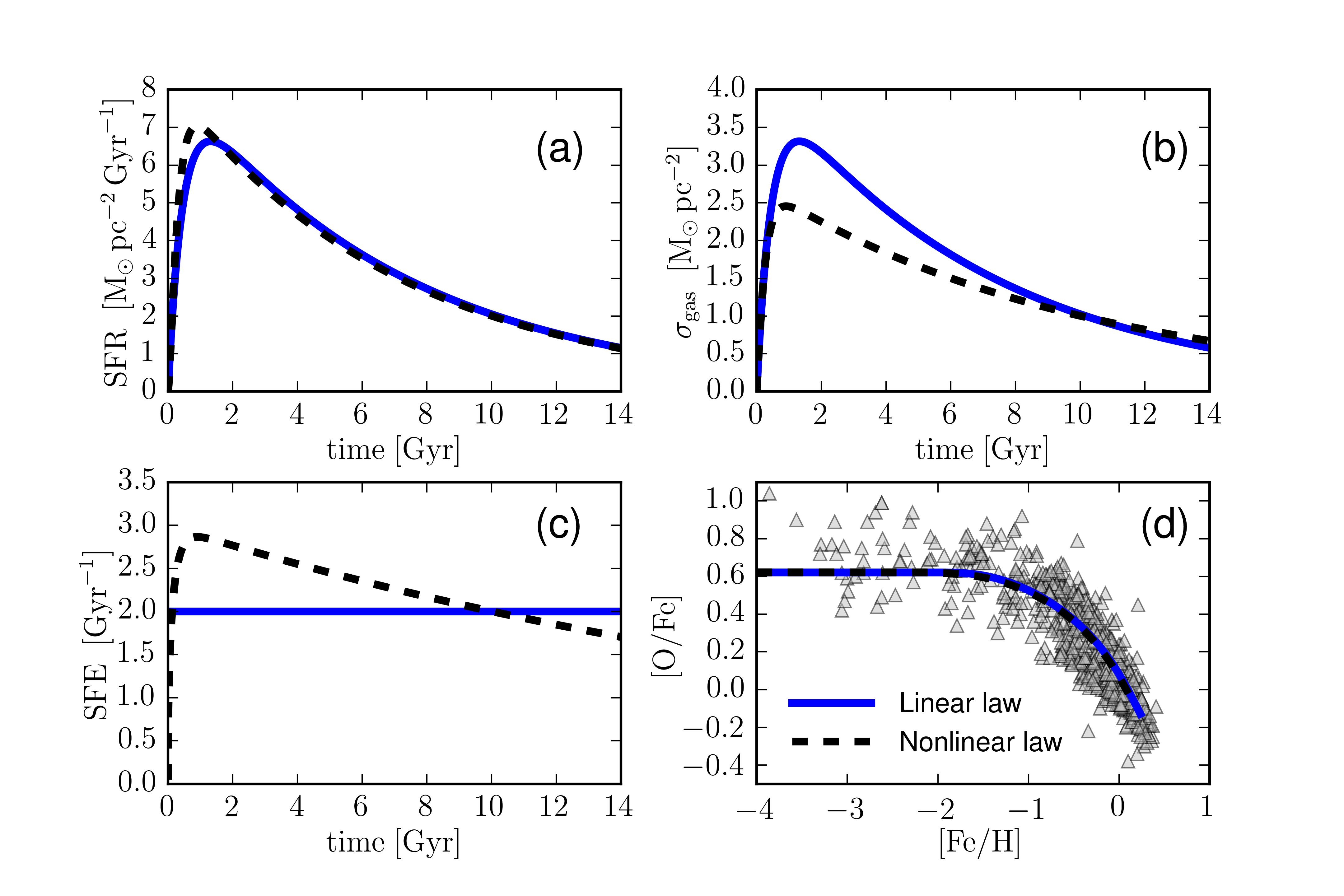}
\caption{ In this Figure, we show the comparison between models with different prescriptions 
for the galaxy SFR. We examine a linear Schmidt-Kennicutt law (solid curves in blue; 
see equation \ref{eq:linear_SFR}) and a nonlinear Schmidt-Kennicutt law (dashed curves in black; 
see equation \ref{eq:original_kennicutt_law}). In panel (a), we 
show the results for the evolution of the SFR as a function of time; in panel (b), 
we show how $\sigma_{\text{gas}}$ is predicted to evolve with time; in panel (c), 
the temporal evolution of the star formation efficiency, defined as 
$\text{SFE}=\psi(r,t)/\sigma_{\text{gas}}(r,t)$, is drawn; finally, in panel (d), 
we show the evolutionary path of the models in the $[\text{Fe/H}]$-$[\text{O/Fe}]$ 
diagram, with the grey triangles corresponding to the observed set of data for the 
stars in the solar neighborhood (see Section \ref{sec:data}). 
All the models assume fixed values of the key free parameters; in particular, 
$\nu_{\text{L}}=2\,\text{Gyr}^{-1}$ (blue solid curves) and 
$\text{SFR}_{0}=2\,\text{M}_{\sun}\,\text{pc}^{-1}\,\text{Gyr}^{-1}$ (black dashed curves); 
infall time scale $\tau=7\,\text{Gyr}$; mass loading factor $\omega=0.4$, and the 
single degenerate scenario for DTD of Type Ia SNe (see also Section \ref{subsec:free_par} for the 
assumed set of free parameters). 
}
     \label{fig:1}
\end{figure*} 

The data set for the observed $[\text{Fe/H}]$-$[\text{O/Fe}]$ and $[\text{Fe/H}]$-$[\text{Si/Fe}]$ 
relations in the solar neighborhood are shown in Fig. \ref{fig:data_by_author}a and Fig. \ref{fig:data_by_author}b, respectively, 
where the data from different works are drawn with different colors. 
Our set of data include both MW halo stars from \citet[red squares]{gratton2003}, \citet[magenta triangles]{akerman2004}, and 
\citet[blue circles]{cayrel2004}, 
and MW halo, thin and thick disc stars from \citet[black triangles]{bensby2014}. 

The data in Fig. \ref{fig:data_by_author} span a wide metallicity range, 
from $[\text{Fe/H}]\approx-4.0\,\text{dex}$ to $[\text{Fe/H}]\approx0.5\,\text{dex}$; they 
show a continuous trend in the $[\text{O/Fe}]$ and
$[\text{Si/Fe}]$ abundance ratios as functions of the $[\text{Fe/H}]$ abundances, 
although highly scattered. There is an initial, slowly decreasing plateau, followed by 
steep decrease, which occurs at different $[\text{Fe/H}]$ abundances and with different slopes 
when halo and disc stars are considered separately. 
This is particularly evident when looking at the $[\text{Si/Fe}]$ ratios in Fig. \ref{fig:data_by_author}b, where the data of the halo 
stars by \citet[red squares]{gratton2003} decrease with a different slope with respect to the disc data by \citet[black triangles]{bensby2014}. 

In Fig. \ref{fig:data_by_component}, we focus on the data set by \citet{bensby2014} and show how the stars of the 
halo (magenta circles), thick disc (blue triangles) and thin disc (red triangles) are distribute in the 
$[\text{Fe/H}]$-$[\text{O/Fe}]$ (upper panel) and $[\text{Fe/H}]$-$[\text{Si/Fe}]$ 
(bottom panel) diagrams. \citet{bensby2014} were able to assign a membership to the majority of the stars 
in their sample, according to the MW stellar component they likely belong to; 
in particular, \citet{bensby2014} adopted a conservative kinematical criterium to retrieve the membership of the stars 
in their sample (see their appendix A). 
The presence of two well separate sequences between thin and thick disc stars, as suggested by many recent works in the literature 
(see, for example, \citealt{nidever2014,kordopatis2015}) is still not evident by looking at the data set in Fig. \ref{fig:data_by_component}, 
since the scatter is still relatively high. Nevertheless, the work by \citet{bensby2014} claimed that thin and thick disc stars 
potentially exhibit two separate abundance trends,  especially in the range $-0.7 \lesssim \text{[Fe/H]} \lesssim -0.35\,\text{dex}$, 
expected by the authors also in the unpublished GAIA data.

\section{Results: the Milky Way chemical evolution} \label{results} 

\begin{figure*}
\includegraphics[width=0.8\textwidth]{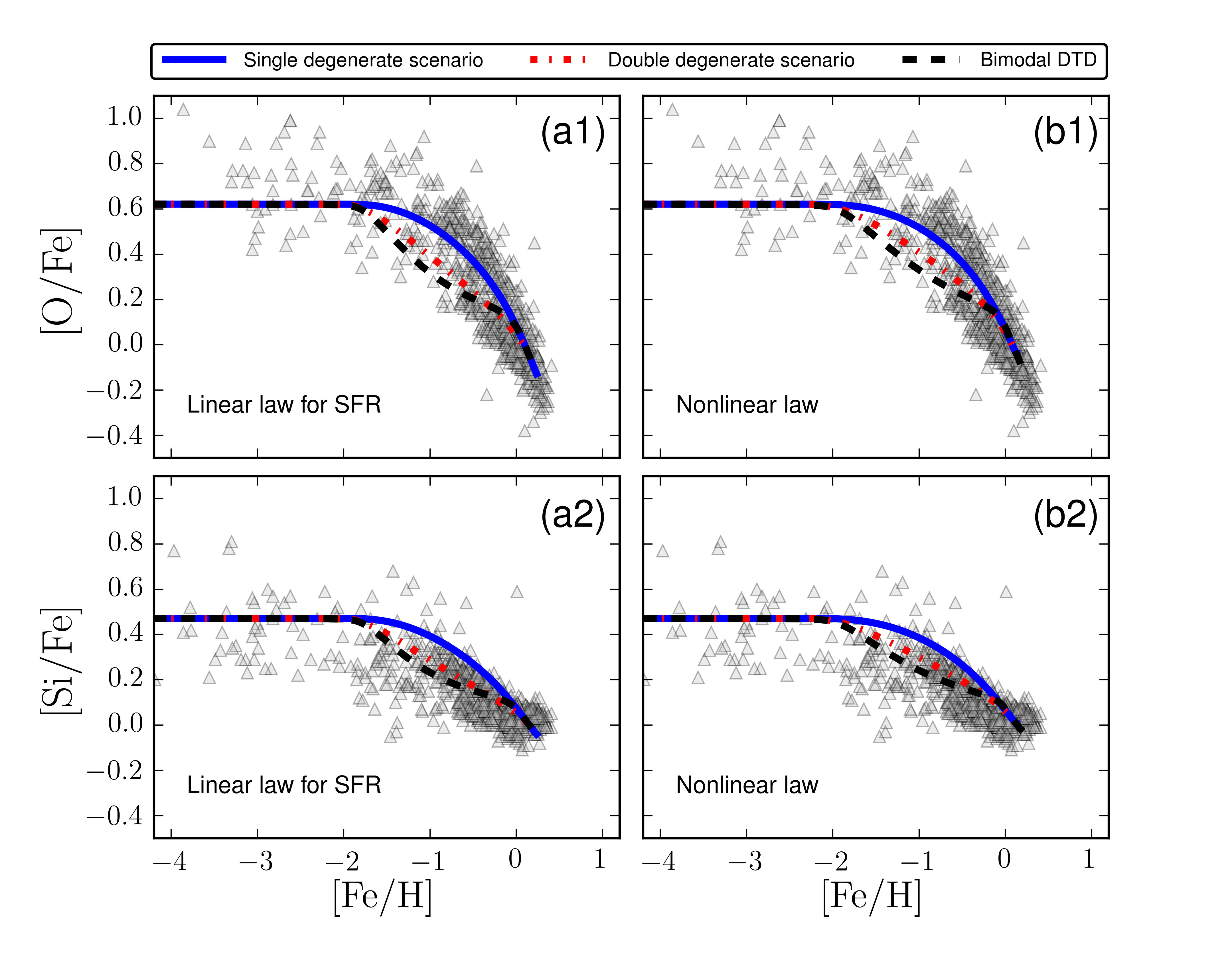}
\caption{ In this Figure, we show the main results of our work for the $[\text{O/Fe}]$ (upper panels) and 
$[\text{Si/Fe}]$ (bottom panels) vs. $[\text{Fe/H}]$ chemical abundance patterns. Curves with different colors 
correspond to different assumptions for the DTD of Type Ia SNe. In this work, we examine the single degenerate 
scenario (solid curves in blue), the double degenerate scenario (dashed-dotted red curves) and the bimodal 
DTD (dashed black curves). The panels (a1) and (a2) on the left show the predictions of the models 
with the linear Schmidt-Kennicutt law (see equation \ref{eq:linear_SFR}), while the panels (b1) and (b2) on the right 
show the models with the nonlinear Schmidt-Kennicutt law (see equation \ref{eq:original_kennicutt_law}). 
}
     \label{fig:2}
\end{figure*} 

In this Section, we present the results of our chemical evolution model, 
which is based on the methods and equations as described in Section \ref{sec:model}. 
We study the MW chemical evolution, 
with the aim of reproducing the observed $[\text{O/Fe}]$ and $[\text{Si/Fe}]$ abundance ratios as functions of 
the $[\text{Fe/H}]$ abundance. 

We firstly explore the effect of different prescriptions for the galaxy SFR. Successively, we investigate 
the effect of assuming different DTDs for Type Ia SNe.

\subsection{Exploring the effect of different laws for the Galaxy SFR} 

In Fig. \ref{fig:1}, we compare the predictions of our best model with the linear Schimidt-Kennicutt law 
(solid lines in blue) with a similar model assuming a nonlinear Schmidt-Kennicutt law. In particular, 
in Fig. \ref{fig:1}a, we show the predicted evolution of the galaxy SFR as a function the galaxy lifetime; in Fig. \ref{fig:1}b, 
the predicted evolution of $\sigma_{\text{gas}}$ with time; in Fig. \ref{fig:1}c, 
we show how the predicted star formation efficiency evolves as a function of time; finally, in Fig. \ref{fig:1}d, we show 
our results for the $[\text{O/Fe}]$  vs. $[\text{Fe/H}]$ chemical abundance pattern.  

By looking at Fig. \ref{fig:1}a, the predicted SFR is very similar when assuming the linear and the nonlinear 
Schmidt-Kennicutt law, with similar absolute values for $\nu_{\text{L}}$ and $\text{SFR}_{0}$. 
The differences are always remarkable in the evolution of $\sigma_{\text{gas}}$ 
with time (see Fig. \ref{fig:1}b), particularly in the earliest stages of the galaxy evolution; in 
fact, the model with the nonlinear Schmidt-Kennicutt law (dashed line in black) 
always predicts higher surface gas mass densities than the model with 
the linear Schmidt-Kennicutt law (solid blue line) for $t\lesssim10\,\text{Gyr}$. 
That can be explained by looking at  Fig. \ref{fig:1}c, 
where we can appreciate that -- for galaxy evolutionary times $t\lesssim10\,\text{Gyr}$ -- the model with the nonlinear law has always higher 
SFEs  (see equation \ref{eq:sfe_nonlinear})
than the model with the linear law, which has a fixed star formation efficiency 
$\nu=2\,\text{Gyr}^{-1}$. 
Since the predicted evolution of the galaxy SFR is very similar when assuming a linear or nonlinear Schmidt-Kennicutt law, 
the differences in the predicted $[\text{O/Fe}]$  vs. $[\text{Fe/H}]$ 
relation are negligible (see Fig. \ref{fig:1}d). 

The predicted evolutionary path of our models in the $[\text{Fe/H}]$-$[\text{O/Fe}]$ diagram in 
Fig. \ref{fig:1}d can be explained as follows, by means of the so-called 
``time delay model'' \citep{tinsley1979,greggio1983,matteucci1986}. 

\begin{enumerate}
\item The plateau at very low $[\text{Fe/H}]$ abundances 
stems from the chemical enrichment by massive stars, dying as core-collapse SNe, for which we assume IRA and 
hence a constant ratio between the net yields of oxygen and iron per stellar generation; in fact, in our simplified model, 
we dot not assume the stellar yields of oxygen and iron per stellar generation to depend upon the 
metallicity (see also \citealt{vincenzo2016}). 
\item The initial plateau is then followed by a decrease, which is due to 
the delayed contribution of Type Ia SNe, injecting large amounts of iron into the galaxy ISM. 
The position of the knee in the $[\text{Fe/H}]$-$[\text{O/Fe}]$ relation 
is determined by the galaxy star formation efficiency. 
In particular, decreasing the star formation efficiency determines a slower 
production of iron and $\alpha$-elements by massive stars and 
hence the decrease of the $[\text{O/Fe}]$ ratios occurs at lower 
$[\text{Fe/H}]$ abundances. 
\end{enumerate}
\noindent If we assumed the original \citet{kennicutt1998} law, which 
prescribes $\text{SFR}_{0,\text{K98}}=2.5\times10^{-1}\,\text{M}_{\sun}\,\text{pc}^{-2}\,\text{Gyr}^{-1}$, 
the star formation efficiencies -- as defined by equation \ref{eq:sfe_nonlinear} -- would be always 
roughly one order of magnitude lower than the findings of our best model, determining a decrease of the $[\text{O/Fe}]$ ratios 
at very low $[\text{Fe/H}]$ abundances, at variance with data.

\subsection{Exploring the effect of different DTDs for Type Ia SNe} 

In Fig. \ref{fig:2}, we show the effect of varying simultaneously the assumed DTDs for Type Ia SNe (curves with different 
colors in all the panels) and the prescriptions for the SFR law (Fig. \ref{fig:2}a1 and Fig. \ref{fig:2}a2 
show the results of the models with the linear Schmidt-Kennicutt law for $[\text{O/Fe}]$ and $[\text{Si/Fe}]$ vs. 
$[\text{Fe/H}]$, respectively, while Fig. \ref{fig:2}b1 and Fig. \ref{fig:2}b2 show the results for the law by \citealt{kennicutt1998}). 

By looking at Fig. \ref{fig:2}, we can appreciate that different distributions of delay times for Type Ia SNe 
determine different behaviors of the models. 
Whatever is the assumed prescription for the galaxy SFR, 
the best agreement with the set of data for $[\text{O/Fe}]$ vs. $[\text{Fe/H}]$ is achieved by the model with the single degenerate 
scenario (blue solid lines), while the predicted trend of the $[\text{Si/Fe}]$ vs. $[\text{Fe/H}]$ with the 
single degenerate scenario is always above the data; such an offset for silicon is due to 
the still large uncertainty in the nucleosynthetic stellar yields of this chemical element. 

The models with the bimodal DTD of \citet[dashed curves in black]{mannucci2006} predict 
a remarkable change in the behavior of the declining trend of the $[\alpha/\text{Fe}]$ ratios, which is 
due to the assumed secondary population of Type Ia SNe, which contributes by $50$ per cent to the global distribution 
of the delay times; 
this tardy component bolsters the iron pollution of the ISM at later times. 
Moreover, the $[\alpha/\text{Fe}]$ ratios with the bimodal DTD decline with the steepest slope among the assumed DTDs 
because of the too large number of prompt Type Ia SNe; this result is in agreement with the findings of 
\citet{matteucci2006,matteucci2009,yates2013}; all these studies agree 
that the average number 
of prompt Type Ia SNe can vary from $\sim15$ per cent to a maximum of $\sim30$ 
per cent with respect to the integrated number of Type Ia SNe, hence a percentage which should be 
lower than the $50$ per cent assumed by \citet{mannucci2006}, although 
these authors concluded that for their data also a percentage of $30$ per cent 
could be acceptable. 
Finally, the models with the double degenerate scenario (dashed-dotted red curves) predict a steeper declining trend 
of the $[\alpha/\text{Fe}]$ ratios than the single degenerate scenario as first Type Ia SNe explode. 

It is worth noting that, if there are separate sequences for thick and thin disc stars, then the double degenerate DTD 
(and even the unlikely bimodal DTD) may provide a better fit to the MW thin disc than the single degenerate DTD, within our simplified model.

\section{Conclusions and discussion} \label{conclusions}

In this work, we have presented a new theoretical framework for following the chemical evolution of galaxies, by 
assuming IRA for chemical elements synthesized and restored by massive stars on short typical time-scales 
and the DTD formalism for the delayed chemical enrichment by Type Ia SNe. The main 
assumptions of our model are the galaxy gas mass assembly history, the stellar yields and the initial mass function. 
Finally, the SFR law represents also an other fundamental phenomenological assumption in the model. 


We have derived a very simple and general formula (equation \ref{laplace}), 
relating the Laplace transform of the 
galaxy SFH with the Laplace transform of the galaxy gas mass assembly history, provided the galaxy SFR 
follows a linear Schmidt-Kennicutt law, namely $\psi(r,t)\propto\sigma_{\text{gas}}(r,t)$. 
This formula can be used both to derive the 
galaxy SFH from the  assumed gas infall law and to solve the corresponding inverse problem (i.e., retrieving the galaxy gas accretion history 
from the observed SFH). 
We remark on the fact that the Laplace transform method for solving ordinary differential equations 
can be used when the latter are linear in the unknown. 

In this work, we have also considered the case of a nonlinear 
ordinary differential equation for the evolution of the galaxy surface gas mass density with time, which can be obtained 
by assuming $\psi(r,t)\propto\sigma_{\text{gas}}(r,t)^{k}$, with $k=1.4$ (nonlinear Schmidt-Kennicutt law, with same same 
index $k$ as given in \citealt{kennicutt1998}). 
In this case, the equation has been solved numerically; then, the derived galaxy SFH has been assumed as an input for the chemical evolution 
model. 

We have shown that the differential equation for the evolution 
of a generic chemical element $X$ can be solved with standard numerical methods (e.g., the Runge-Kutta algorithm), once the 
galaxy SFH has been theoretically determined from our assumptions.  
We have applied our model to reproduce the 
$[\text{O/Fe}]$ and $[\text{Si/Fe}]$ vs. $[\text{Fe/H}]$ chemical abundance patterns as observed in the solar neighborhood, 
by exploring the effect of different DTDs for Type Ia SNe 
and different prescriptions for the SFR law. 
We have assumed the Galaxy disc 
to assemble by means of a single 
gas accretion episode, with typical time scale $\tau=7\,\text{Gyr}$. 
In any case, our method can be easily 
extended also for a two-infall model \citep{chiappini1997}. 

Our model with the single degenerate scenario for Type Ia SNe 
provides a very good agreement with the 
observed $[\text{Fe/H}]$-$[\text{O/Fe}]$ relation in the MW. 
Since the nucleosynthetic stellar yields of the other $\alpha$-elements, like Si, Ca or Mg, still suffer 
of large uncertainty, the agreement between model and data for these chemical elements is still not good. 
We also conclude that, if there are two separate sequences between thin and thick disc data, then our models with
 the double degenerate scenario (or even the bimodal DTD) may provide a better fit to the thin disc data. 

We remark on the fact that a linear relation between the 
SFR and $\sigma_{\text{gas}}$ seems not to be the best fit to data, 
unless the star formation efficiency is defined as the inverse of the typical dynamical time scale 
of the system. Nevertheless, we have shown that our models with a nonlinear Schmidt-Kennicutt law 
provide very similar results for the galaxy SFH and chemical abundance patterns as the model with the linear 
Schmidt-Kennicutt law. 

 
We are aware that the assumption of a constant yield per stellar generation is a strong 
approximation, determining the predicted flat trend of the $[\alpha/\text{Fe}]$ ratios at low $[\text{Fe/H}]$ abundances, 
at variance with data showing a remarkable scatter. 
This scatter can be reproduced only by introducing an element of stochasticity in the model 
(for example, in the stellar yields, in the IMF or in the SFR). 
Moreover, the $[\alpha/\text{Fe}]$ ratios at very low $[\text{Fe/H}]$ abundances show a global trend, 
which can be reproduced only by relaxing IRA, and hence by including stellar lifetimes and variable nucleosynthetic stellar yields of 
massive stars, as in the detailed numerical codes of chemical evolution.
As Type Ia SNe start exploding, the $[\alpha/\text{Fe}]$ ratios steeply decrease, in excellent agreement with data. 

The theoretical method, as presented in this work, differs from previous works in the literature, which 
similarly recover the 
galaxy chemical evolution from the assumed SFH 
(see, for example, \citealt{erb2006,homma2015,weinberg2016}), because we initially start from 
the assumption of a galaxy gas mass assembly history. Then, the galaxy SFH has been retrieved either 
by means of equation (\ref{laplace}), which makes use of the Laplace transform to simplify 
the solution of the differential equation for the evolution of the galaxy gas mass, or by means 
of numerical techniques. 
Our theoretical framework can be 
generalized for 
any choice of the galaxy gas infall law. 

Our method for solving chemical evolution of galaxies 
can be easily included in other complementary stellar population synthesis models, 
by taking into account chemical elements -- like iron -- 
which are restored with a time delay from the star formation event. In this way, one can easily decouple the 
evolution of iron and oxygen, which contribute to the total metallicity of the galaxy ISM with 
different relative fractions as a function of the galaxy lifetime.



\section*{Acknowledgements}

We thank S. Recchi and F. Belfiore for their remarks upon reading the manuscript and for insightful discussions; 
we also thank two anonymous referees for their suggestions and constructive comments. 
FM and ES acknowledge funding from PRIN-MIUR~2010-2011 project 
`The Chemical and Dynamical Evolution of the Milky Way and Local Group Galaxies', prot.~2010LY5N2T.  


\bsp

\label{lastpage}

\end{document}